\def\rfr#1{eq. (\ref{#1})}

\def\dert#1#2{\frac{{{d}}{#1}}{{{d}}{#2}}}              

\def\virg#1{``#1''}

\def\bb#1#2#3{\bibitem[\protect\citeauthoryear{#1}{#2}]{#3}}
\def\eqi{\begin{equation}}
\def\eqf{\end{equation}}
\def\eqia{\begin{eqnarray}}
\def\eqfa{\end{eqnarray}}
\def\rp#1#2{{#1\over#2}} \def\lb#1{\label{#1}}

\def\bds#1{\boldsymbol{#1}}

\documentclass{aastex}
\usepackage{w-greek}
\usepackage{url}
\usepackage{amsmath,amsthm,amscd,amssymb}
\usepackage{latexsym}
\usepackage{graphicx,epsfig}

\RequirePackage{color}

\newcommand{\emaila}{lorenzo.iorio@libero.it}

\begin{document}

\title{Constraints on the location of a putative distant massive body in the Solar System  from recent planetary data}
\shortauthors{L. Iorio}

\author{Lorenzo Iorio\altaffilmark{1} }
\affil{Ministero dell'Istruzione, dell'Universit\`{a} e della Ricerca (M.I.U.R.)-Istruzione. Fellow of the Royal Astronomical Society (F.R.A.S.). International Institute for Theoretical Physics and
Advanced Mathematics Einstein-Galilei. Permanent address for correspondence: Viale Unit\`{a} di Italia 68, 70125, Bari (BA), Italy.}

\email{\emaila}

\begin{abstract}
We analytically work out the long-term variations caused on the motion of a planet orbiting a star by a very distant, pointlike massive object X. Apart from the semi-major axis $a$, all the other Keplerian osculating orbital elements  experience long-term variations which are complicated functions of the orbital configurations of both the planet itself and of X.  We infer constraints on the minimum distance $d_{\rm X}$ at which  X may exist by comparing our prediction of the long-term variation of the longitude of the perihelion $\varpi$ to the latest empirical determinations of the corrections $\Delta\dot\varpi$ to the standard Newtonian/Einsteinian secular precessions of several solar system planets  recently estimated by independent teams of astronomers.
  We obtain the following approximate lower bounds on  $d_{\rm X}$ for the assumed masses of X quoted in brackets: $150-200$ au (Mars), $250-450$ au ($0.7\  m_{\oplus}$),  $3500-4500$ au ($4\ m_{\rm Jup}$).
\end{abstract}

\keywords{Planets $\cdot$ Planet X $\cdot$ Solar System perturbations}

\section{Introduction}
Does not any other planet of the Sun exist in addition to those already known?
Have not the remote peripheries of the solar system  saved any further big surprises for us? It is a long time \citep{Low,Pick1,Pick2,Schue49} that such  questions$-$still well alive \citep{Lyka,TOAJ,Mat2,Fer011}$-$are well rooted in
the astronomical community, having often resurfaced in different contexts and with changed forms;. For a recent, popular review, see \citet{Schillig}. Here we limit ourselves to recall that
\begin{itemize}
\item
The hypothesis of a stellar-like companion object (Nemesis) orbiting the Sun was postulated for explaining terrestrial extinction periodicity \citep{Raup}, thought to be mediated by comet showers \citep{Whit,Davis,Hills,Hut,Van93,Mull02}. A recent critical review about the evidence for and against astronomical impacts on climate change and mass extinctions can be found in \citet{Bail09}; see  \citet{Mel}  for recent developments of the investigations on the role of Nemesis in such phenomena.
\item Anomalies in the  distribution of the aphelia and the orbital elements of  the comets resident in the outer region of the Oort cloud led to postulating the existence of a Sun-bound Jovian mass body \citep{Mat1}. Other studies on the interplay among a putative Planet X and comets are, e.g., \citet{Mat0,Mur99,Horn02}. For  recent, quantitative investigations on such a Jupiter-type object, now named
     Tyche \citep{Mat2} and not to be confused with Nemesis since it would not be able to induce comet storms, see \citet{Mat2}. Another recent analysis can be found in \citet{Fer011}.
\item The existence of an as yet undiscovered planet orbiting at trans-Plutonian distances was postulated \citep{Mara97,Coll00,Bru02,Mel03,Mel04,Mat05,Gom06,Lyka}, with changeable fortune, to explain several features pertaining the dynamical history and the architecture of the Trans-Neptunian Belt. See \citet{Lyka}  for recent developments of such a scenario.
\item A putative Planet X was also invoked \citep{Gun70,Raw70,Seidel71,vanF81,vanF82,And87,Seidel88,Har88,Gom88,Gom89,Pow89,vanF91} to explain certain seeming irregularities in the orbital motions of Uranus and Neptune which showed up in ancient data \citep{Bru92}; such an issue was later settled by \citet{Sta93} with a  re-analysis of the observations by including more recent radiometric points from the Voyager 2 automatic spacecraft which allowed for a more accurate determination of the mass of Neptune.
    See \citet{Sta96} for issues related with the orbital residuals of Pluto, likely of non-dynamical origin.
    \item \citet{Harr77} suggested that, as an explanation of the peculiar properties of certain pulsars with anomalously small period derivatives, the barycenter of the solar system is accelerated, possibly because of a hitherto undetected companion star of the Sun in a bound or open orbit. See also \citet{Cow83} and \citet{Zaka} for further studies on such a topic.
        \item \citet{TOAJ} found that a distant, pointlike massive object in the outer regions of the solar system may explain the anomalous retrograde precession\footnote{Anyway, more recent data analyses \citep{Pit010} yield values for such an anomaly which are statistically compatible with zero. See Section \ref{periprece}.} of the perihelion of Saturn preliminarily estimated \citep{Pit} by analyzing some radiotechnical data points from the Cassini spacecraft exploring the Saturnian system.
\end{itemize}
Here by means of the expression \virg{Planet X} we broadly  denote a pointlike object, having a mass  approximately as large as that of Mars or  larger, up to typical values of brown dwarfs ($m_{\rm X}\approx 80\ m_{\rm Jup}$) or even red dwarfs ($m_{\rm X}\approx 0.5\ {\rm M}_{\odot}$) according to certain scenarios, located at large distances from the Sun, and for which no direct observational evidence, based on the detection of electromagnetic radiation of different wavelengths emitted or reflected by it in natural processes, is (yet?) available.

We also note that investigations on Planet X have direct connections with fundamental physics as well since it has been shown \citep{MOND1,MOND2} that the action of a distant body located towards the Galactic center is dynamically equivalent to that of the External Field Effect (EFE) in the planetary regions of the solar system within the framework of the Modified Newtonian Dynamics (MOND). Moreover, \citet{Foot}  considered the possible existence of solar system's planets made of  non-annihilating mirror matter, which is one of the candidates for the non-baryonic Dark Matter which is believed to constitute about  $22\%$ of the matter content of the Universe \citep{DM}. For other researches on the existence of mirror matter in the solar system, see \citet{Foot2}; a general overview on the mirror matter scenario and its observational implications can be found in, e.g., \cite{Foot02}. For earlier studies on such a topic, see \citet{Blin,Khlo1,Khlo2}.
Finally, the subtle perturbing effects due to a putative X may be important in realistically assessing the error budget in several high-precision tests of standard post-Newtonian gravity proposed or to be performed in the solar system arena \citep{Iorio2}.

In this paper, we  use recent and accurate observational determinations of the motion of some planets of the solar system to put tighter constraints on the location of Planet X with respect to those existing in literature. The plan of the paper is as follows. In Section \ref{due} we  analytically work out the effects that a distant body X would cause on the orbital motion of a closer and (relatively) fast-moving planet. More specifically, we adopt the Lagrange perturbative scheme to calculate, without simplifying approximations concerning the orbital configurations of both the  planet and of X, the long-term variations of all the Keplerian orbital elements of the perturbed planet. In Section \ref{periprece} we turn to the latest observational determinations of the planetary motions. In particular, we compare our theoretical prediction for the long-term variation of the perihelion with the most recent determinations of the corrections to the standard precessions of the perihelia of the major bodies of the solar system obtained  by independent teams of astronomers with the latest ephemerides. Our goal is to look at the minimum distance at which Planet X may exist as a function of its position in the sky.  In Section \ref{dete} we review the past and future sky surveys and compare our bounds on the minimum distance of X to their obtained and expected results.  Section \ref{conclu} is devoted to the conclusions.
\section{Analytical calculation}\lb{due}
Let us consider a remote, point-like  object X of mass $m_{\rm X}$ and located at distance $r_{\rm X}$ from a central body of mass $M$.
 Its action   on a closer test particle orbiting $M$ at distance $r$  can be modelled as due to the following quadrupolar potential \citep{Hogg}, accurate up to terms of order $\mathcal{O}(r^2/r_{\rm X}^2)$,
\eqi U_{\rm X} = \rp{\mathcal{K}_{\rm X}}{2}\left[r^2 -3\left(\bds r\bds\cdot\bds{\hat{l}} \right)^2\right],\lb{ux}\eqf
where
\eqi \mathcal{K}_{\rm X}\doteq \rp{Gm_{\rm X}}{r_{\rm X}^3}\eqf is the tidal parameter of X, while $\bds{\hat{l}}=\left\{l_x,l_y,l_z\right\}$ is a unit vector directed towards X determining its position in the sky. We purposely will not adopt any specific parameterization for it in order to make our results as general as possible.
In \rfr{ux} $\bds r=\left\{x,y,z\right\}$ refers to the perturbed test particle.

In order to compute the effects of \rfr{ux} on the orbital motion of a test particle we shall use standard perturbative techniques.
The use the eccentric anomaly $E$ \citep{Murr} as fast variable of integration  turns out to be more convenient to make the forthcoming integration more tractable.
The average of \rfr{ux} over one orbital revolution of the perturbed test particle is
\begin{equation}
\left\langle U_{\rm X} \right\rangle  \doteq \left(\rp{n}{2\pi}\right)\int_0^{P_{\rm b}} U_{\rm X}\ dt =  \rp{{\mathcal{K}_{\rm X}}a^2}{32}\mathcal{U}\left(I,\Omega,\omega; \bds{\hat{l}}\right),\lb{us}
\end{equation}
with
\begin{equation}
\begin{array}{lll}
\mathcal{U} &\doteq & - \left(2 + 3 e^2\right)  \left( -8  + 9 l_x^2 + 9 l_y^2 +
        6 l_z^2\right)   -120 e^2 \sin 2\omega \left(l_x \cos\Omega + l_y \sin\Omega\right)  \left[l_z \sin I +\right. \\ \\
    &+&\left.\cos I \left(l_y \cos\Omega - l_x \sin\Omega\right) \right]  - 15 e^2 \cos 2\omega \left[3
 \left(l_x^2 - l_y^2\right)  \cos 2\Omega + 2 \left(
              l_x^2 + l_y^2 - 2
                    l_z^2\right)  \sin^2 I -\right. \\ \\
                     &-&\left. 4 l_z \sin 2I \left(l_y
\cos\Omega - l_x \sin\Omega\right)  + 6 l_x l_y \sin 2\Omega\right]  -
          6 \left(2 + 3 e^2\right)  \left[ \left(l_x^2 - l_y^2\right)   \cos 2\Omega \sin^2 I +\right. \\ \\
           &+&\left. 2 l_z \sin 2I \left(l_y
                    \cos\Omega - l_x \sin\Omega\right)  +
                2 l_x l_y \sin^2 I \sin 2\Omega\right]  - 3\cos 2I \left\{ \left(2 + 3
                    e^2\right)  \left(l_x^2 + l_y^2 - 2 l_z^2\right)  +\right. \\ \\
                     &+&\left. 5 e^2 \cos 2\omega \left[ \left(l_x^2 - l_y^2\right)   \cos 2\Omega + 2 l_x l_y \sin 2\Omega\right] \right\}.
\end{array}\lb{uavera}
\end{equation}
In \rfr{us} and \rfr{uavera}    $a,e,I,\Omega,\omega$ are the semi-major axis, the eccentricity, the inclination of the test particle's orbit to the reference $\{X,Y\}$ plane, the longitude of the ascending node and the argument of pericenter, respectively; $n\doteq\sqrt{GM/a^3}$ is the unperturbed Keplerian mean motion connected to the unperturbed orbital period $P_{\rm b}$ by $n=2\pi/P_{\rm b}$.
Note that \rfr{us} and \rfr{uavera} are exact: neither approximations in $e$ nor in $I$ were used.
In the integration  $\bds{\hat{l}}$ was kept fixed over one orbital revolution of the perturbed test particle, as it is reasonable given the assumed large distance of X with respect to it.

The standard Lagrange planetary equations \citep{BeFa},
in which  the disturbing function, i.e.  $\left\langle U_{\rm X}\right\rangle$ in our case, appears,
can be used to straightforwardly derive the long-term variations of all the standard six Keplerian osculating orbital elements of the perturbed particle. We  also considered the longitude of the pericenter $\varpi\doteq\Omega+\omega$ since its secular precession is one of the parameters usually estimated by the astronomers when they fit the dynamical force models of their ephemerides to huge planetary data records \citep{Pit05,Pit,Pit010,Fie010,FiengaJournees010}.

From  the Lagrange planetary equation for the semi-major axis \citep{BeFa}, it can be noted that  $a$ does not change since \rfr{uavera} does not contain the mean anomaly $\mathcal{M}$. Instead, all the other Keplerian orbital elements experience non-vanishing long-term variations. They are
\begin{itemize}
\item
\eqi\dert e t = \rp{15\mathcal{K}_{\rm X}e\sqrt{1-e^2}}{16n}\mathcal{E}\left(I,\Omega,\omega; \bds{\hat{l}}\right),\lb{eccecazzo}\eqf
with
 \begin{equation}
\begin{array}{lll}
\mathcal{E} & \doteq & -8  l_z \cos 2\omega \sin I \left(l_x \cos\Omega + l_y \sin\Omega\right)  +
    4 \cos I \cos 2 \omega \left[ -2  l_x l_y \cos 2\Omega +\right.         \\ \\
    &+&\left.  \left(l_x^2 - l_y^2\right)  \sin 2\Omega\right]  +
    \sin 2\omega \left[ \left(l_x^2 - l_y^2\right)   \left(3 +
                \cos 2I\right)  \cos 2\Omega +
          2 \left(l_x^2 + l_y^2 - 2 l_z^2\right)  \sin^2 I -\right. \\ \\
          &-&\left.
          4 l_z \sin 2I \left(l_y \cos\Omega -
                l_x \sin\Omega\right)  +
          2 l_x l_y \left(3 + \cos 2I\right)  \sin 2\Omega\right].
          \end{array}
\end{equation}
\item
\eqi\dert I t = \rp{3\mathcal{K}_{\rm X}}{4\sqrt{1-e^2}n}\mathcal{I}\left(I,\Omega,\omega;\bds{\hat{l}}\right),\lb{inclicazzo}\eqf
in which
\begin{equation}
\begin{array}{lll}
\mathcal{I} & \doteq &
\left[l_z \cos  I +
        \sin  I \left( -l_y  \cos \Omega +
              l_x \sin \Omega\right)   \right]  \left\{5 e^2 l_z \sin  I \sin  2\omega + \right. \\ \\
              &+& \left. 5 e^2 \cos  I \sin  2\omega \left(l_y \cos \Omega -
              l_x \sin \Omega\right)  +  \left[2 + e^2\left(3  +
              5 \cos  2\omega\right)\right]  \left(l_x \cos \Omega +
              l_y \sin \Omega\right)   \right\}.
\end{array}
\end{equation}
\item
\eqi\dert\Omega t = -\rp{\mathcal{K}_{\rm X}}{4n\sqrt{1-e^2}}\mathcal{N}\left(I,\Omega,\omega;\bds{\hat{l}}\right),\lb{nodocazzo}\eqf
with
\begin{equation}
\begin{array}{lll}
\mathcal{N} & \doteq & 3  \csc I \left[l_z \cos I + \sin I \left( -l_y  \cos\Omega +
l_x \sin\Omega\right) \right]  \left\{ -2  l_z \sin I +\right. \\ \\
 &+&\left.\cos I \left[ -2  +e^2\left( -3  + 5  \cos 2\omega\right)\right]  \left(l_y \cos\Omega -
       l_x \sin\Omega\right)  + \right. \\ \\
 &+&\left.      e^2 \left[l_z \left( -3  + 5 \cos 2\omega\right)  \sin I - 5 \sin 2\omega \left(l_x \cos\Omega + l_y \sin\Omega\right) \right] \right\}.
\end{array}
\end{equation}
\item
\eqi \dert\omega t = -\rp{3\mathcal{K}_{\rm X}}{16 n\sqrt{1-e^2}}\mathcal{P}\left(I,\Omega,\omega;\bds{\hat{l}}\right)\lb{perizoma},\eqf
with
\begin{equation}
\begin{array}{lll}
\mathcal{P} & \doteq & -\left(1 - e^2\right)\left(-8 + 9 l_x^2 + 9 l_y^2 + 6 l_z^2\right) + \left(l_x^2+l_y^2 - 2 l_z^2\right)
\left[-2-3e^2 -5 \cos 2 I +\right. \\ \\
&+&\left.5\cos 2\omega\left( -1+2e^2 + \cos 2I\right)\right] + \cos2\Omega
\left\{ \left(l_x^2-l_y^2\right) \left[ -1+6e^2 +\right.\right. \\ \\
&+& \left.\left. 5\left(2e^2-3\right)\cos 2\omega +10\cos 2I\sin^2\omega\right] -20\left(2-e^2\right)l_x l_y\cos I\sin2 \omega \right\}+\\ \\
&+& 2l_z\cos\Omega\left\{  2l_y\cot I  \left[\left(1-e^2\right)\left(-3+5\cos 2\omega\right)   +10\cos 2I\sin^2\omega\right]  -\right.\\ \\
&-&\left. 5l_x
\left[2-3e^2 -\left(2-e^2\right)\cos 2I\right] \csc I\sin 2\omega \right\} - l_z\csc I\sin\Omega
\left\{ 4l_x\cos I \left[\left(1-\right.\right.\right. \\ \\
&-&\left.\left.\left. e^2\right)\left(-3 + 5\cos 2\omega\right)  +10\cos 2I\sin^2\omega\right] + 10l_y\sin 2\omega\left[ 2-3e^2 -\right.\right. \\ \\
&-&\left.\left.\left.\left(2-e^2\right)\cos 2 I\right] \right\} +2\sin 2\Omega\left[-5\left(3-2e^2\right)l_x l_y \cos 2\omega  + l_x l_y \left(-1 + 6 e^2  + \right.\right.\right. \\ \\
&+&\left.\left. 10 \cos 2 I\sin^2\omega\right) +5 \left(2-e^2\right)\left(l_x^2-l_y^2\right)\cos I \sin 2\omega\right].
\end{array}
\end{equation}
\item
\eqi \dert\varpi t =  \frac{\mathcal{K}_{\rm X}}{128 n\sqrt{1-e^2}}\left[\mathcal{G}\left(I,\Omega,\omega;\bds{\hat{l}}\right)+\mathcal{H}\left(I,\Omega,\omega;\bds{\hat{l}}\right)\right],\lb{pericazzo}\eqf
where
\begin{equation}
\begin{array}{lll}
-\rp{\mathcal{G}}{24\left(1-e^2\right)} & \doteq &
8 - 9 l_x^2 - 9 l_y^2 - 6 l_z^2 -
          10 \left(l_x^2 + l_y^2 - 2 l_z^2\right)  \cos 2\omega \sin^2 I +\\ \\
          &+&
          8 l_z \cos\Omega \sin I \left[l_y \cos I \left( -3  + 5 \cos 2\omega\right)  -
                5 l_x \sin 2\omega\right]  +\\ \\
          &+&\cos 2\Omega \left[ -3  \left(l_x^2 - l^2_y\right)   \left(5 \cos 2\omega + 2 \sin^2 I\right)  - 40 l_x l_y \cos I \sin 2\omega\right]  -\\ \\
          &-& 8 l_z \sin I \left[l_x \cos I \left( -3  + 5 \cos 2\omega\right)  +
                5 l_y \sin 2\omega\right]  \sin\Omega +\\ \\
                &+&
          2 \left[ -3  l_x l_y \left(5 \cos 2\omega +
                      2 \sin^2 I\right)  +
                10 \left(l_x^2 - l_y^2\right)   \cos I \sin 2\omega\right]  \sin 2\Omega -\\ \\
                &-&
          \cos 2I \left\{3 \left(l_x^2 + l_y^2 - 2 l_z^2\right)  +
                5 \cos 2\omega \left[ \left(l_x^2 - l_y^2\right)   \cos 2\Omega +
                      2 l_x l_y \sin 2\Omega\right] \right\}   ,
\end{array}
\end{equation}
and
\begin{equation}
\begin{array}{lll}
\mathcal{H} & \doteq & -96 \left[l_z \cos I +
          \sin I \left( -l_y  \cos\Omega +
                l_x \sin\Omega\right) \right]  \left\{ -2  l_z \sin I +\right. \\ \\
          &+&\left.\cos I \left[ -2  +e^2\left(- 3  +
                5  \cos 2\omega\right)\right]  \left(l_y \cos\Omega -
                l_x \sin\Omega\right)  +\right. \\ \\
          &+&\left. e^2 \left[l_z \left( -3  + 5 \cos 2\omega\right)  \sin I - 5 \sin 2\omega \left(l_x \cos\Omega +
                      l_y \sin\Omega\right) \right] \right\} \tan\left(I/2\right).
\end{array}
\end{equation}
\item
\eqi \dert{\mathcal{M}} t = -\rp{\mathcal{K}_{\rm X}}{16 n}\mathcal{A}\left(I,\Omega,\omega; \bds{\hat{l}}\right), \lb{anocazzo}\eqf
with
\begin{equation}
\begin{array}{lll}
\mathcal{A}&\doteq & \left(7 + 3 e^2\right)  \left( -8  + 9 l_x^2 + 9 l_y^2 +
          6 l_z^2\right)  +
     120 \left(1 + e^2\right)  \sin  2\omega\left(l_x \cos \Omega +\right. \\ \\
     &+&\left. l_y \sin \Omega\right)  \left[l_z \sin  I +
                \cos  I \left(l_y \cos \Omega -
                      l_x \sin \Omega\right) \right]  +
          15 \left(1 + e^2\right)  \cos  2\omega\left[3 \left(l_x^2 - \right.\right.\\ \\
          &-&\left.\left.
          l_y^2\right)  \cos  2\Omega +
                2 \left(l_x^2 + l_y^2 - 2 l_z^2\right)  \sin^2  I -
                4 l_z \sin  2I \left(l_y \cos \Omega -
                      l_x \sin \Omega\right)  +\right. \\ \\
                      &+&\left.
                6 l_x l_y \sin  2\Omega\right]  +
          6 \left(7 +3 e^2\right)  \left[ \left(l_x^2 - l_y^2\right)  \cos  2\Omega \sin^2  I +
                2 l_z \sin  2I \left(l_y \cos \Omega -\right.\right. \\ \\
                &-&\left.\left.
                      l_x \sin \Omega\right)  +
                2 l_x l_y \sin ^2 I \sin  2\Omega\right]  +
          3\cos  2I \left\{ \left(7 + 3 e^2\right)  \left(l_x^2 + l_y^2 -
                      2 l_z^2\right)  +\right. \\ \\
                      &+&\left.
                5 \left(1 + e^2\right)  \cos  2\omega\left[ \left(l_x^2 - l_y^2\right)  \cos  2\Omega +
                      2 l_x l_y \sin  2\Omega\right] \right\}   .
\end{array}
\end{equation}
\end{itemize}
It is not possible to simplify such formulas by $a-priori$ orienting one of the reference axes along $\bds{\hat{l}}$ because the position of X is not known. In the case of the solar system, the reference $\{X,Y\}$ plane is typically the mean ecliptic at the epoch J$2000.0$, with the $X$ axis directed towards the mean equinox at the same epoch. Our calculation holds for any orientation of the reference frame.

Other calculations, performed in the framework of the lunisolar perturbations on artificial planetary satellites, exist in literature \citep{Koz,Cook,Gurf}. They differ from ours in the choice of the orbital elements investigated, the calculational techniques adopted, and the representation of the unit vector of the disturbing body. A straightforward comparison with our results is, thus, not possible. \citet{Koz}, by using a specific parameterization for the geocentric lunar unit vector $\bds{\hat{l}}$ in terms of some orbital elements of it, adopted the Lagrange planetary equations for $a,e,I,\Omega,\omega$. Anyway, he did not average the disturbing function over one orbital revolution of the satellite, retaining only the terms in which its mean longitude does not appear. Moreover, \citet{Koz} did not release explicit expressions for the variations of $e,I$ at all, while those for $\Omega$ and $\omega$ are presumably incomplete and difficult to evaluate since it is unclear how they were obtained. \citet{Cook} used the more cumbersome Gauss\footnote{He dubbed them as Lagrange's planetary equations.} equations for the variations of the osculating orbital elements to compute the long-term variations of $a,e,I,\Omega,\omega+\Omega\cos I$ by means of the true anomaly $f$ as fast variable of integration. Also \citet{Cook} used a particular representation of the geocentric $\bds{\hat{l}}$, expressed  in terms of its Keplerian orbital elements.
\citet{Gurf}, working in the framework of a perturbed Martian satellite,  employed the satellite's mean anomaly with the Lagrange planetary equations in the integration over one orbital revolution for $\dot a,\dot e,\dot I,\dot \Omega,\dot \omega$. They also adopted two specific parameterizations for $\bds{\hat{l}}$ related to different choices for the reference $\{X,Y\}$ plane.

  Our results  can be used not only in view of the fact that estimated corrections $\Delta\dot\varpi$ of the standard Newtonian/Einsteinian precessions of the longitudes of the perihelia are nowadays available for all the eight planets and Pluto \citep{Pit010,Fie010,FiengaJournees010}, but also because it seems that \textcolor{black}{also the nodes  will receive their due attention by the astronomers} in the near future  \citep{FiengaJournees010}. \textcolor{black}{The first determination of the corrections to their standard secular precession for the first six planets can be found in \citet{Fie011}: we propose to use them in further analyses}. Moreover, \rfr{eccecazzo}-\rfr{anocazzo} can also be useful in  several\footnote{See http://exoplanet.eu/ on the WEB.} exoplanetary scenarios. Indeed,   in many cases the discovered extrasolar planets have distant companions \citep{How010}, or their existence is postulated  to explain certain observed features of the orbital motions \citep{WASP33b}.

\section{Confrontation with the observations}\lb{periprece}
In Table \ref{perirate} we quote the latest determinations of the corrections $\Delta\dot\varpi$ to the standard Newtonian/Einsteinian  secular precessions of the longitudes of the perihelia of all the major bodies of the solar system estimated by various authors with some of the most recent ephemerides.
\begin{table*}[ht!]
\caption{Estimated corrections $\Delta\dot\varpi$, in milliarcseconds per century (mas cty$^{-1}$), of the standard Newtonian/Einsteinian secular precessions of the longitudes of the perihelia $\varpi$ of the eight planets and Pluto determined with the EPM2008 \protect{\citep{Pit010}}, the INPOP08 \protect{\citep{Fie010}},  and the INPOP10a \protect{\citep{FiengaJournees010}} ephemerides. Concerning the values quoted in the third column from the left, they are those corresponding to the smallest uncertainties reported in \protect{\citep{Fie010}}. Note the small uncertainty in the correction to the precession of the terrestrial perihelion, obtained by processing Jupiter VLBI data \protect{\citep{Fie010}}.
}\label{perirate}
\centering
\bigskip
\begin{tabular}{llll}
\hline\noalign{\smallskip}
Planet & $\Delta\dot\varpi$  \protect{\citep{Pit010}}  & $\Delta\dot\varpi$  \protect{\citep{Fie010}} &  $\Delta\dot\varpi$  \protect{\citep{FiengaJournees010}} \\
\noalign{\smallskip}\hline\noalign{\smallskip}
Mercury & $ -4 \pm 5 $  & $ -10\pm 30$ & $ 0.2\pm 3$ \\
Venus & $ 24\pm 33$  & $-4\pm 6 $ & $ - $ \\
Earth & $ 6\pm 7$  & $ 0 \pm 0.016 $ & $ - $\\
Mars & $ -7\pm 7$  & $0\pm 0.2 $ & $ - $\\
Jupiter & $ 67\pm 93$  & $142\pm 156$ & $ - $\\
Saturn & $ -10\pm 15$ & $-10\pm 8$ & $ 0\pm 2$ \\
Uranus & $ -3890\pm 3900$  & $0\pm 20000$ & $ - $\\
Neptune & $ -4440\pm 5400 $  & $0\pm 20000$ & $ - $\\
Pluto & $ 2840 \pm 4510 $  & $-$ & $ - $\\
\noalign{\smallskip}\hline\noalign{\smallskip}
\end{tabular}
\end{table*}
The effect of Planet X was not explicitly included in the dynamical force models fitted to the observations, so that, in principle, the extra-rates of the perihelia in Table \ref{perirate} account for it.
Here we shall compare our theoretical prediction of \rfr{pericazzo} to the corrections of Table \ref{perirate} in order to infer upper bounds on
$\mathcal{K}_{\rm X}$, so to
constrain the minimum distance $d_{\rm X}$ at which X may be located for different values of its mass $m_{\rm X}$.

By expressing the unit vector $\bds{\hat{l}}$ of X in terms of its ecliptic latitude $\beta_{\rm X}$ and longitude $\lambda_{\rm X}$
%
%
%
%
%
%
and by equating $\dot\varpi$ of \rfr{pericazzo} to $\Delta\dot\varpi$ of Table \ref{perirate} for each planet, it is possible to plot the minimum distance $d_{\rm X}$ of X as a function of $\beta_{\rm X},\lambda_{\rm X}$ for given values of its putative mass $m_{\rm X}$.
In Figure \ref{grossi} we plot such bounding surfaces obtained from the accurate extra-precessions of the  perihelia of the Earth, Mars and Saturn  \citep{Fie010,FiengaJournees010} for  $m_{\rm X}=4\  m_{\rm Jup}$, corresponding to Tyche \citep{Mat2}, $m_{\rm X}=0.7\  m_{\oplus}$, considered in the scenario by \citet{Lyka}, and $m_{\rm X}=m_{\rm Mars}$.
\begin{figure*}[ht!]
\centering
\begin{tabular}{ccc}
\epsfig{file=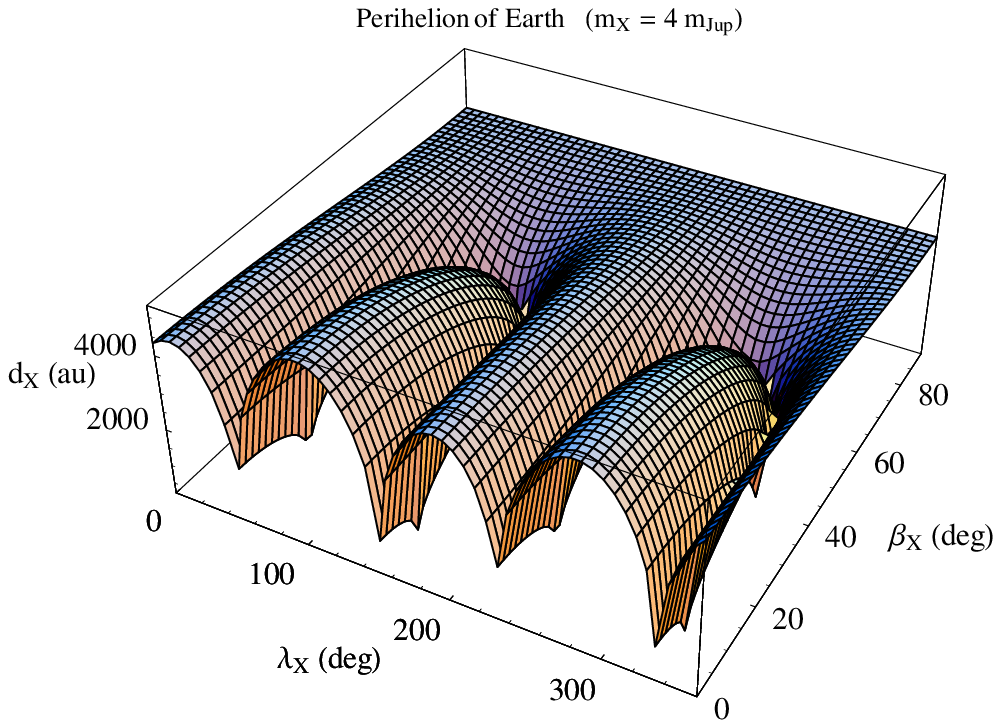,width=0.25\linewidth,clip=} & \epsfig{file=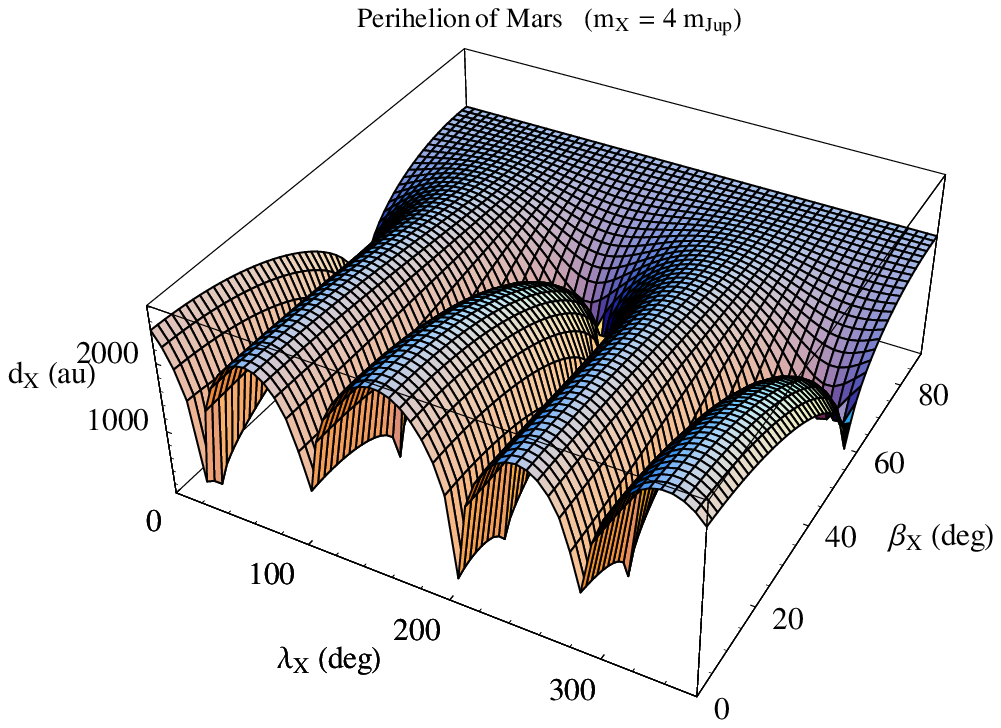,width=0.25\linewidth,clip=} &
\epsfig{file=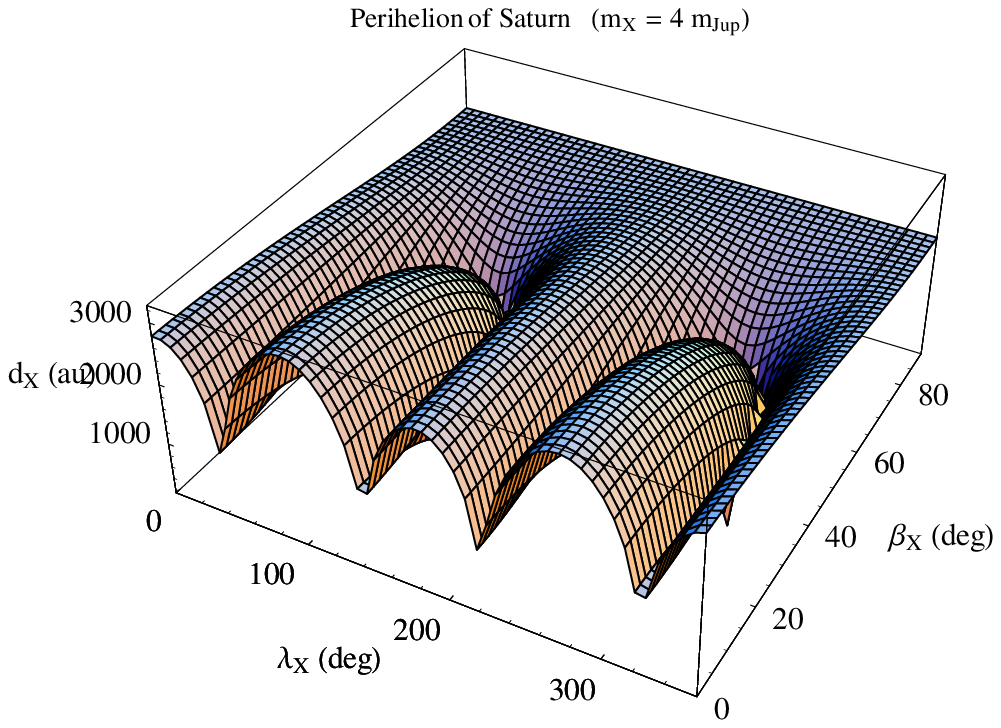,width=0.25\linewidth,clip=} \\
\epsfig{file=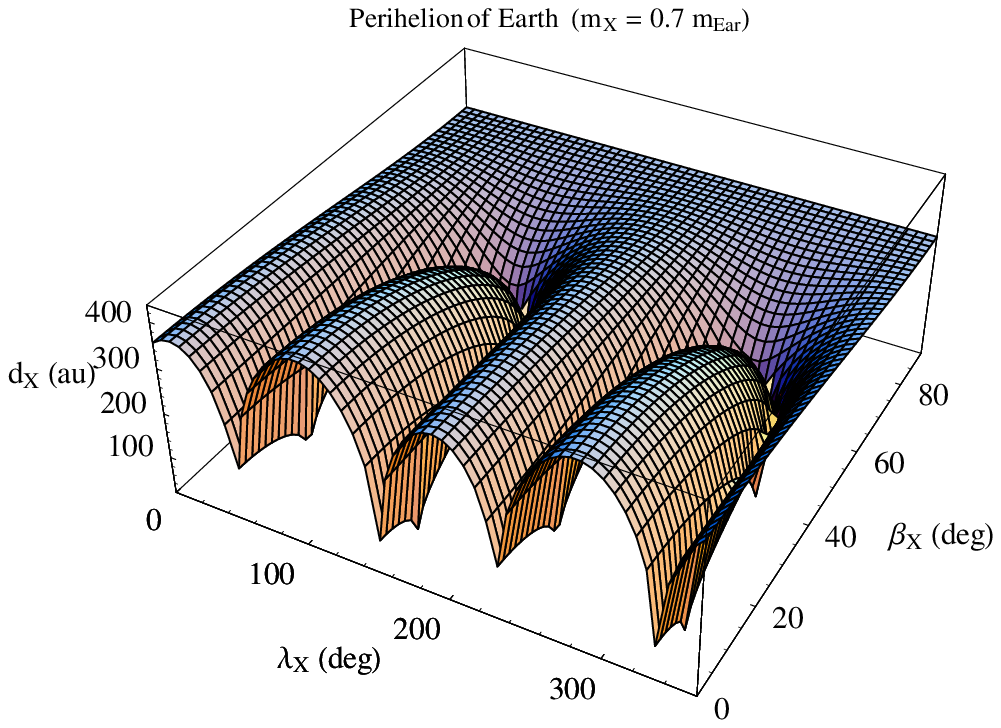,width=0.25\linewidth,clip=} & \epsfig{file=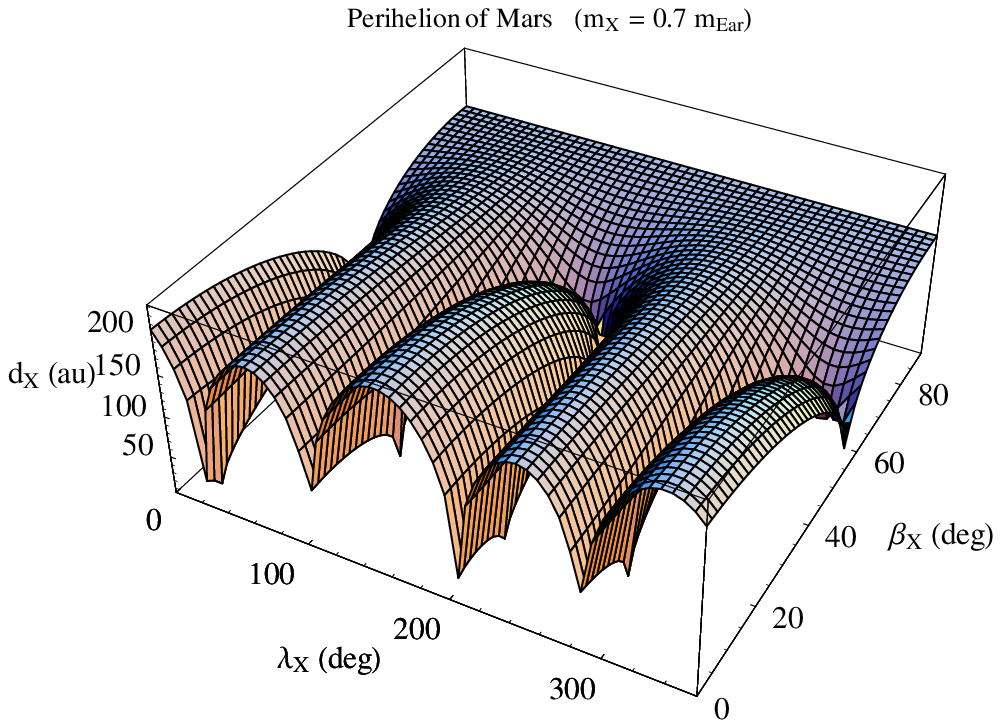,width=0.25\linewidth,clip=} &
\epsfig{file=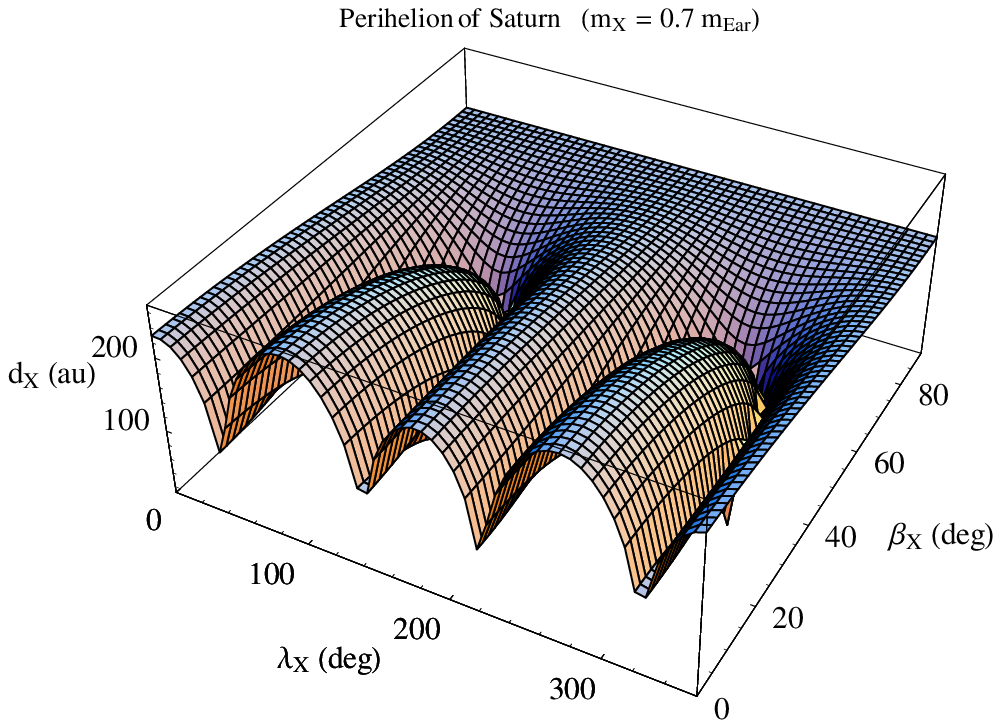,width=0.25\linewidth,clip=}\\
\epsfig{file=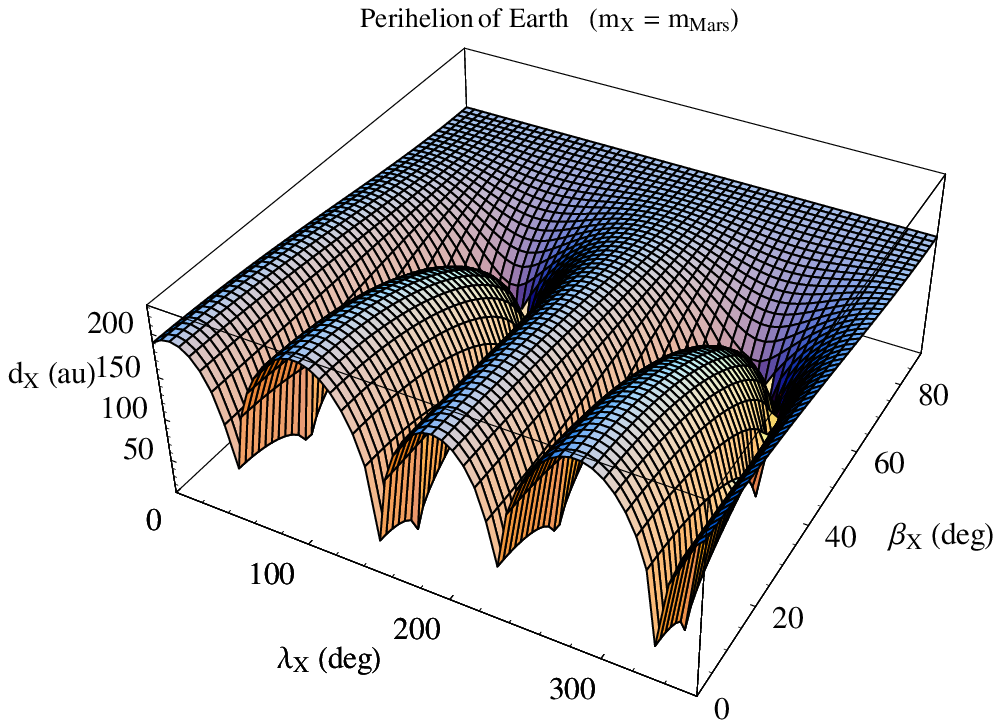,width=0.25\linewidth,clip=} & \epsfig{file=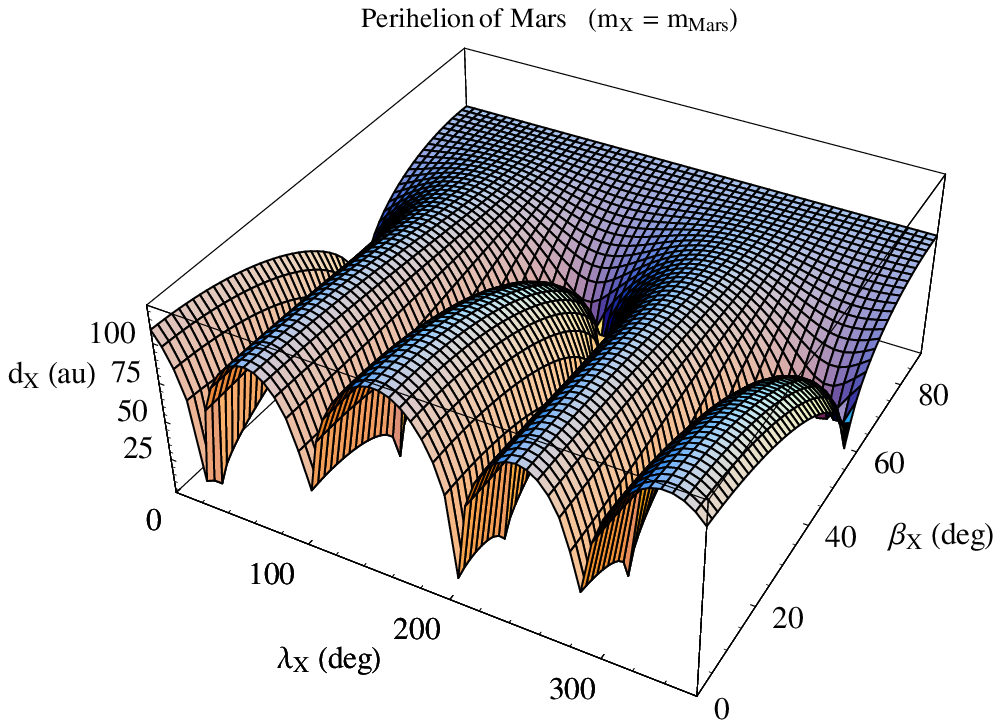,width=0.25\linewidth,clip=} &
\epsfig{file=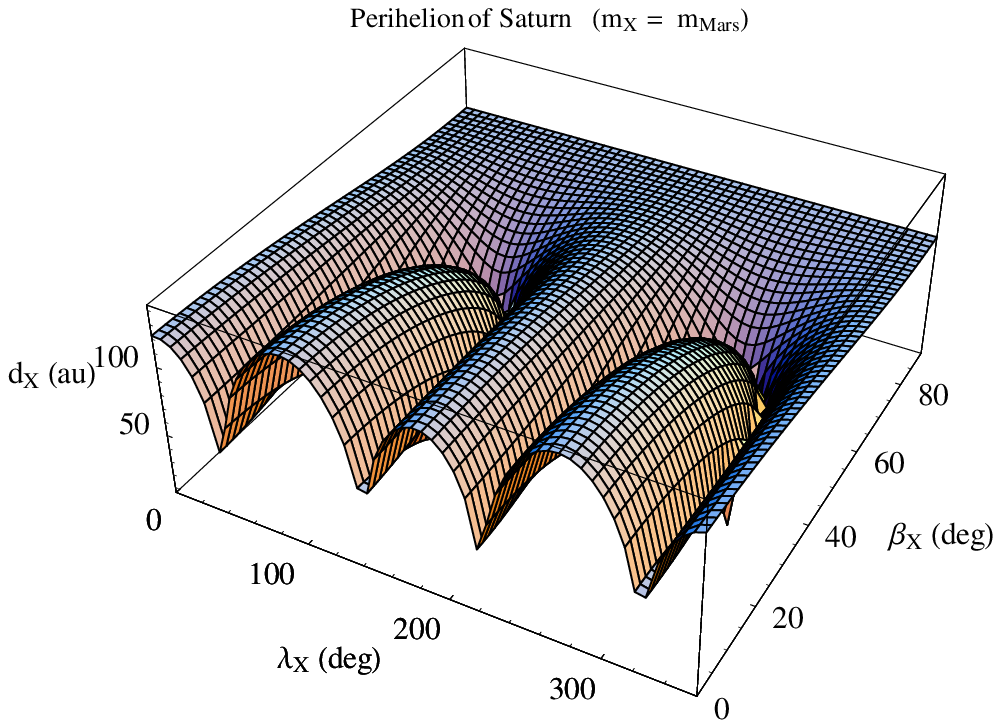,width=0.25\linewidth,clip=}
\end{tabular}
\caption{Constraints on the minimum distance $d_{\rm X}$, in au, for   $m_{\rm X}=4\  m_{\rm Jup}$ (first row from the top), $m_{\rm X}= 0.7\  m_{\oplus}$ (second row from the top) and $m_{\rm X}= m_{\rm Mars}$ (third row from the top) as a function of the ecliptic latitude $\beta_{\rm X}$ and longitude $\lambda_{\rm X}$ of X. The perihelia of the Earth (left column), Mars (middle column) and Saturn (right column) were used according to  $\Delta\dot\varpi_{\oplus} = 0\pm 0.016$ mas cty$^{-1}$  and $\Delta\dot\varpi_{\rm Mars} = 0\pm 0.2$ mas cty$^{-1}$ \protect{\citep{Fie010}}, and $\Delta\dot\varpi_{\rm Sat} = 0\pm 2$ mas cty$^{-1}$ \protect{\citep{FiengaJournees010}}.}\lb{grossi}
\end{figure*}
It turns out that the bounds on $d_{\rm X}$ at high ecliptic latitudes are rather independent of $\lambda_{\rm X}$.
For each of the perihelia there are a few sky locations, mainly close to the ecliptic,  in which the dynamically inferred minimum distances of X are  unrealistically small. Anyway, such positions differ from each other in such a way that a compensation occurs. Thus, a Tyche-like body cannot be located at less than about $3500-4500$ au, an Earth-sized mass should not exist at less than about $250-450$ au, while a rocky planet as big as Mars cannot be closer than approximately $150-200$ au. The bounding surfaces by the Saturnian perihelion are generally smoother, with a milder excursion among the extremum values. They  are rather close to those determined by the perihelion of the Earth, whose extra-precession is 2 orders of magnitude more accurate than for Saturn (cfr. Table \ref{perirate}). Future improvements in the accuracy in determining the orbit of the ringed planet from continuous tracking of Cassini will straightforwardly yield more stringent constraints.
It is possible to infer $d_{\rm X}$ also for values of $m_{\rm X}$ different from those  used in Figure \ref{grossi}, dubbed $\overline{m}_{\rm X}$,  by scaling the corresponding lower bounds $\overline{d}_{\rm X}$ by the multiplicative factor $\xi_{\rm X}\doteq (m_{\rm X}/\overline{m}_{\rm X})^{1/3}$.

Our bounds are  tighter than those obtained by \citet{Iorio09} with the correction to the precession of the perihelion of Mars estimated by \citet{Pit05}.
\section{Possibilities of direct detections}\label{dete}
Concerning the ability of several performed, ongoing or planned wide area sky surveys to directly reveal a putative Planet X from its emitted or reflected electromagnetic radiation of different wavelengths, let us recall the following.

Limiting ourselves to optical wavelengths only, bodies as large as Jupiter or Neptune would have apparent visual magnitudes of about 20 and 23, respectively, at 1000 au, so that the largest telescopes may image them like just very dim dots \citep{Fer011}. At 10000 au it would impossible to detect them with any optical telescope available today \citep{Fer011}. Actually, our tightest constraints from the Earth's perihelion  pose a jovian-sized object at not less than approximately $3000-4000$ au; the minimum allowed distance for a Neptune-like mass would be of no more than about $1110$ au. Since low-temperature bodies emit more efficiently in the infrared, surveys operating at such wavelengths are more suitable to look for Planet X, especially those based in space because of the atmospheric extinction in the infrared.

All the wide area sky surveys performed so far mostly explored regions close to the ecliptic.
The ecliptic, all-sky optical survey by \citep{Tom61} concluded that the minimum distance at which an Earth-sized body could be located is 81 au; similar findings were reached by the nearly ecliptic optical survey by \citet{Kow89}. Our dynamically inferred lower bounds are about $3-6$ times more stringent. The Spacewatch optical survey \citep{Lar07}, conducted within 10 deg from the ecliptic, was sensitive to Mars-sized objects out to 300 au and Jupiter-sized planets out to 1200 au; its negative findings, especially with respect to the detection of a Jupiter-like body, are well explained by our analysis since its minimum distance dynamically inferred from the planetary perihelia is just more than 1000 au for those ecliptic latitudes.
According to \citet{Mat2}, Planet X would have to be greater than $7-10$ $m_{\rm Jup}$ and
closer than $6000-25000$ au for a possible detection in the infrared, ground-based  Two Micron All Sky Survey (2MASS) \citep{2mass}. Our minimum distance for a body with $m_{\rm X}=7\ m_{\rm Jup}$ may be of the order of $1300$ au in a few locations near the ecliptic, and 4800 au for high latitudes.
\citet{Mat2} note also that the negative  search results of the InfraRed Astronomical Satellite (IRAS) survey\footnote{Certain  rumors appeared in 1983 \citep{otoole} in mass-media about IRAS and Planet X,  indirectly caused by \citet{rumors}. Clarifications in specialistic literature, not spread by mass-media, appeared later \citep{clari1,clari2}.} \citep{iras}  suggest that an object with $m_{\rm X}=2-5\ m_{\rm Jup}$ must have
a current minimum distance $d_{\rm X}=2000-10000$ au, respectively; our dynamically inferred bounds for the minimum allowed distance of such a X are
in the range $\approx 3000-4500$ au for most of the positions in the sky.
The all-sky synoptic Tycho-2 optical survey excluded the presence of a main-sequence star
above the hydrogen-burning limit within 1 pc$= 2.06\times 10^5$ au \citep{hog}.

Moving to ongoing or forthcoming projects,
the all-sky synoptic survey by the Panoramic Survey Telescope and Rapid Response System (Pan-STARRS) \citep{Jew03} would be able to reveal massive planets such as Neptune
not beyond about 800 au,
while a body with $m_{\rm X} = 0.1\ M_{\odot}$ would be undetectable for
$r_{\rm X}>2000$ au. Conversely, our dynamically inferred lower bounds for such kind of bodies in the ecliptic are up to $\approx 1100$ au and $\approx 13400$  au, respectively.
A Jupiter-sized
planet at 2000 au over the whole sky could, in principle, be detected by GAIA \citep{gaia} with astrometric microlensing; actually, our lower limits for the admissible distance for such a body are globally more stringent. Also the putative existence of a X with $m_{\rm X}=3\ m_{\rm Jup}$ at 20000 au could be put on the test \citep{gaia}.
Possible planets in the outer regions ($>$1000 au) of the solar system may be revealed by the mesolensing technique \citep{meso}.
Another future all-sky survey which may be useful in detecting distant objects far from the ecliptic is the Large Synoptic Survey Telescope (LSST) \citep{lsst}.
\citet{Babi} proposed to use observations of the Cosmic Microwave Background (CMB)
to place constraints on the mass, distance, and size distribution of small objects in the Kuiper Belt and inner Oort Cloud. \citet{Planck} envisaged the possibility of using the data from the ongoing Planck mission to detect the thermal emission of pointlike objects of the solar system.
The recently launched Wide-field Infrared Survey Explorer (WISE) \citep{wise}
 should be able to detect  a Neptune-sized object  out to 700 au, which is, in fact, a distance smaller than our dynamically inferred bounds on $d_{\rm X}$ for such a body. Instead, a Jupiter-mass object would be detectable out to 1 light year$=63239.7$ au, where it would still be within the Sun's zone of gravitational control. A larger object of $2–3$ Jupiter masses would be visible at a distance of up to 7 to 10 light years \citep{WISE2}, i.e. $4.4-6.3\times 10^5$ au.
Actually, WISE has recently discovered its first ultra-cold brown dwarf, named WISEPC J045853.90+643451.9, at about $6-10$ pc \citep{Mainz} corresponding to $1.2-2.0\times 10^6$ au.
\section{Summary and conclusions}\lb{conclu}
We analytically computed the long-term variations of all the Keplerian osculating orbital elements of a test particle orbiting a primary and acted upon by a pointlike, distant object X. We assumed that it is located at a much larger distance $r_{\rm X}$  than the perturbed  particle, so that the spatial position of X was kept fixed in the integration over one  orbital revolution of the particle. With a perturbative, first-order calculation accurate up to terms of order $\mathcal{O}(r^2/r^2_{\rm X})$ we found  that all the Keplerian osculating orbital elements, apart from its semi-major axis $a$, undergo long-term variations  depending  on the orbital geometries of both the perturbed and the perturbing objects in a complicated way.

We compared our analytical prediction for $\dot\varpi$  with the latest empirically determined corrections $\Delta\dot\varpi$ to the standard Newtonian/Einsteinian secular precessions of the perihelia of several solar system planets estimated by different teams of astronomers. We  inferred constraints on the minimum distance $d_{\rm X}$ at which  X may be located  as a function of its ecliptic latitude $\beta_{\rm X}$ and longitude $\lambda_{\rm X}$ for different values of its mass $m_{\rm X}$. We adopted   the masses of  Mars, $0.7\  m_{\oplus}$ and $4\  m_{\rm Jup}$. The use of the perihelia is effective in constraining $d_{\rm X}$ for most of the sky locations: the availability of $\Delta\dot\varpi$ for more than one planet allowed to mutually overlap the forbidden  regions, thus further strengthening the lower bounds on $r_{\rm X}$ also in those positions of the sky in which the use of a single planet at a time would, instead, yield weaker constraints. The most stringent bounds came from the extra-rates of the perihelia of Saturn, Mars, and, especially, the Earth thanks to the inclusion of the Jupiter VLBI data.
Future improvements for Saturn, due to continuous ranging to Cassini,  should yield tighter constraints.

The minimum distance at which a remote body X with  $70\%$ of the mass of the Earth can exist is of the order of $250-450$ au. For other values of the mass of X, we have   $150-200$ au (Mars), and $3500-4500$ ($4\ m_{\rm Jup}$). Such lower bounds, which are independent of any speculation concerning the physical properties of X because they are solely based on its gravitational perturbations exerted on known planets,  are  tighter than those previously obtained by us in literature. Moreover, they are useful  to better interpreting the results of past surveys aimed to directly detect X, and also to clarify what could realistically be expected from ongoing and future observational campaigns.

\textcolor{black}{We intend to further extend our analysis in future by using the newly determined corrections to the standard secular precessions of the planetary perihelia and nodes as well by \citet{Fie011}.}

We remark that the corrections to the perihelion precessions used here were independently obtained by various astronomers without explicitly modeling the dynamical action of X itself. Thus, it might, in principle, have been somewhat \virg{absorbed} and partially removed in the usual process of  estimation of the solution's parameters like, e.g., the initial state vectors of the planets, especially if its magnitude was very small. To circumvent such an issue one should, actually, re-process the entire observational data set with modified dynamical force models explicitly including X itself, and solving for one or more dedicated parameters as well. Then, it would, e.g., be possible to look at the consequences of the inclusion of X on the values of the other usually estimated parameters.



\end{document}